\documentclass[twocolumn,tighten,twocolappendix]{aastex701}
\usepackage{graphicx}
\usepackage{amsmath}
\usepackage{amsfonts}
\usepackage{amssymb}
\usepackage{xcolor}

\newcommand{\sunrise}{{\sc{Sunrise~iii}}}
\begin{document}

\title{Origin of small-scale evaporation flows deep in the chromosphere during a solar flare} 

\author[0000-0002-9270-6785]{L. P. Chitta}
\affiliation{Max-Planck-Institut für Sonnensystemforschung, Justus-von-Liebig-Weg 3, 37077 Göttingen, Germany}
\email[show]{chitta@mps.mpg.de}  

\author[orcid=0000-0003-3490-6532,sname='Smitha']{H.~N.~Smitha} \affiliation{Max-Planck-Institut für Sonnensystemforschung, Justus-von-Liebig-Weg 3, 37077 Göttingen, Germany}\email{narayanamurthy@mps.mpg.de}


\author[orcid=0000-0003-1409-1145,sname='Iglesias']{Francisco~A.~Iglesias} \affiliation{Max-Planck-Institut für Sonnensystemforschung, Justus-von-Liebig-Weg 3, 37077 Göttingen, Germany}\affiliation{Grupo de Estudios en Heliofísica de Mendoza, CONICET, Universidad de Mendoza, Boulogne sur Mer 683, 5500 Mendoza, Argentina}\email{iglesias@mps.mpg.de}
\author[orcid=0000-0001-6317-4380,sname='Riethmüller']{Tino~L.~Riethmüller} \affiliation{Max-Planck-Institut für Sonnensystemforschung, Justus-von-Liebig-Weg 3, 37077 Göttingen, Germany}\email{riethmueller@mps.mpg.de}
\author[orcid=0009-0009-4425-599X,sname='Feller']{Alex~Feller} \affiliation{Max-Planck-Institut für Sonnensystemforschung, Justus-von-Liebig-Weg 3, 37077 Göttingen, Germany}\email{feller@mps.mpg.de}

\author[orcid=0000-0001-5279-3266,sname='Chen']{Wei~Chen} \affiliation{Key Laboratory of Dark Matter and Space Astronomy, Purple Mountain Observatory, Chinese Academy of Sciences,
Nanjing 210034, China}\email{w.chen@pmo.ac.cn}

\author[orcid=0000-0003-1459-7074,sname='Lagg']{Andreas~Lagg} \affiliation{Max-Planck-Institut für Sonnensystemforschung, Justus-von-Liebig-Weg 3, 37077 Göttingen, Germany}\email{lagg@mps.mpg.de}

\author[orcid=0000-0002-9972-9840,sname='Gandorfer']{Achim~Gandorfer} \affiliation{Max-Planck-Institut für Sonnensystemforschung, Justus-von-Liebig-Weg 3, 37077 Göttingen, Germany}\email{gandorfer@mps.mpg.de}

\author[orcid=0000-0001-6029-7529,sname='Hölken']{Johannes~Hölken} \affiliation{Max-Planck-Institut für Sonnensystemforschung, Justus-von-Liebig-Weg 3, 37077 Göttingen, Germany}\email{hoelken@mps.mpg.de}

\author[orcid=0000-0002-3418-8449,sname='Solanki']{Sami~K.~Solanki} \affiliation{Max-Planck-Institut für Sonnensystemforschung, Justus-von-Liebig-Weg 3, 37077 Göttingen, Germany}\email{solanki@mps.mpg.de}

\author[orcid=0000-0002-3387-026X,sname='del~Toro~Iniesta']{Jose~Carlos~del~Toro~Iniesta} \affiliation{Instituto de Astrofísica de Andalucía, CSIC, Glorieta de la Astronomía s/n, 18008 Granada, Spain}\affiliation{Spanish Space Solar Physics Consortium}\email{jti@iaa.es}
\author[orcid=0000-0002-5054-8782,sname='Katsukawa']{Yukio~Katsukawa} \affiliation{National Astronomical Observatory of Japan, 2-21-1 Osawa, Mitaka, Tokyo 181-8588, Japan}\affiliation{Department of Astronomy, The University of Tokyo, 7-3-1, Hongo, Bunkyo-ku, Tokyo 113-0033, Japan}\affiliation{Department of Astronomical Science, The Graduate University for Advanced Studies (SOKENDAI), 2-21-1 Osawa, Mitaka, Tokyo 181-8588, Japan}\email{yukio.katsukawa@nao.ac.jp}
\author[orcid=0000-0002-0787-8954,sname='Bernasconi']{Pietro~Bernasconi} \affiliation{Johns Hopkins University Applied Physics Laboratory, 11100 Johns Hopkins Road, Laurel, Maryland, USA}\email{pietro.bernasconi@jhuapl.edu}
\author[sname='Berkefeld']{Thomas~Berkefeld} \affiliation{Institut für Sonnenphysik (KIS), Georges-Köhler-Allee 401a, 79110 Freiburg, Germany}\email{thomas.berkefeld@leibniz-kis.de}

\author[orcid=0000-0001-9228-3412,sname='Álvarez-Herrero']{Alberto~Álvarez-Herrero} \affiliation{Instituto Nacional de T\'ecnica Aeroespacial (INTA), Ctra. de Ajalvir, km. 4, E-28850 Torrejón de Ardoz, Spain}\affiliation{Spanish Space Solar Physics Consortium}\email{alvareza@inta.es}
\author[orcid=0000-0001-5616-2808,sname='Kubo']{Masahito~Kubo} \affiliation{National Astronomical Observatory of Japan, 2-21-1 Osawa, Mitaka, Tokyo 181-8588, Japan}\affiliation{Department of Astronomical Science, The Graduate University for Advanced Studies (SOKENDAI), 2-21-1 Osawa, Mitaka, Tokyo 181-8588, Japan}\email{masahito.kubo@nao.ac.jp}
\author[orcid=0000-0001-8829-1938,sname='Orozco~Suárez']{David~Orozco~Suárez} \affiliation{Instituto de Astrofísica de Andalucía, CSIC, Glorieta de la Astronomía s/n, 18008 Granada, Spain}\affiliation{Spanish Space Solar Physics Consortium}\email{orozco@iaa.es}
\author[sname='Grauf']{Bianca~Grauf} \affiliation{Max-Planck-Institut für Sonnensystemforschung, Justus-von-Liebig-Weg 3, 37077 Göttingen, Germany}\email{grauf@mps.mpg.de}
\author[sname='Carpenter']{Michael~Carpenter} \affiliation{Johns Hopkins University Applied Physics Laboratory, 11100 Johns Hopkins Road, Laurel, Maryland, USA}\email{michael.carpenter@jhuapl.edu}
\author[sname='Bell']{Alexander~Bell} \affiliation{Institut für Sonnenphysik (KIS), Georges-Köhler-Allee 401a, 79110 Freiburg, Germany}\email{albe@leibniz-kis.de}
\author[orcid=0000-0001-7764-6895,sname='Martínez~Pillet']{Valentín~Martínez~Pillet} \affiliation{Instituto de Astrofísica de Canarias, Vía Láctea, s/n, E-38205 La Laguna, Spain}\affiliation{Departamento de Astrof\'\i sica, Universidad de La Laguna, E-38206 La Laguna, Spain}\affiliation{Spanish Space Solar Physics Consortium}\email{vmpillet@iac.es}

\author[orcid=0000-0002-7318-3536,sname='Bailén']{Francisco~Javier~Bailén} \affiliation{Instituto de Astrofísica de Andalucía, CSIC, Glorieta de la Astronomía s/n, 18008 Granada, Spain}\affiliation{Spanish Space Solar Physics Consortium}\email{fbailen@iaa.es}
\author[orcid=0000-0002-2055-441X,sname='Blanco~Rodríguez']{Julián~Blanco~Rodríguez} \affiliation{Universitat de Valencia Catedrático José Beltrán 2, E-46980 Paterna-Valencia, Spain}\affiliation{Spanish Space Solar Physics Consortium}\email{julian.blanco@uv.es}
\author[orcid=0000-0003-4319-2009,sname='Castellanos~Durán']{Juan~Sebastián~Castellanos~Durán} \affiliation{Max-Planck-Institut für Sonnensystemforschung, Justus-von-Liebig-Weg 3, 37077 Göttingen, Germany}\email{castellanos@mps.mpg.de}
\author[orcid=0009-0002-6808-5154,sname='Harnes']{Edvarda~Harnes} \affiliation{Max-Planck-Institut für Sonnensystemforschung, Justus-von-Liebig-Weg 3, 37077 Göttingen, Germany}\email{harnes@mps.mpg.de}
\author[orcid=0000-0002-4669-5376,sname='Ishikawa']{Ryohtaroh~T.~Ishikawa} \affiliation{National Institute for Fusion Science, 322-6 Oroshi-cho, Toki City 509-5292, Japan}\email{ishikawa.ryohtaro@nifs.ac.jp}
\author[orcid=0000-0001-7452-0656,sname='Kawabata']{Yusuke~Kawabata} \affiliation{National Astronomical Observatory of Japan, 2-21-1 Osawa, Mitaka, Tokyo 181-8588, Japan}\email{kawabata.yusuke@nao.ac.jp}
\author[orcid=0000-0002-1043-9944,sname='Matsumoto']{Takuma~Matsumoto} \affiliation{Centre for Integrated Data Science, Institute for Space-Earth Environmental Research, Nagoya University, Furocho, Chikusa-ku, Nagoya, Aichi 464-8601, Japan}\email{takuma.matsumoto@gmail.com}
\author[orcid=0000-0002-7044-6281,sname='Oba']{Takayoshi~Oba} \affiliation{Advanced Research Center for Space Science and Technology, Institute of Science and Engineering, Kanazawa University, Kakuma-machi, Kanazawa, Ishikawa 920-1192, Japan}\affiliation{Max-Planck-Institut für Sonnensystemforschung, Justus-von-Liebig-Weg 3, 37077 Göttingen, Germany}\email{oba@mps.mpg.de}
\author[orcid=0000-0003-0175-6232,sname='Siu-Tapia']{Azaymi~L.~Siu-Tapia} \affiliation{Instituto de Astrofísica de Andalucía, CSIC, Glorieta de la Astronomía s/n, 18008 Granada, Spain}\affiliation{Spanish Space Solar Physics Consortium}\email{siu@iaa.es}
\author[orcid=0000-0003-1483-4535,sname='Strecker']{Hanna~Strecker} \affiliation{Instituto de Astrofísica de Andalucía, CSIC, Glorieta de la Astronomía s/n, 18008 Granada, Spain}\affiliation{Spanish Space Solar Physics Consortium}\email{streckerh@iaa.es}
\author[orcid=0000-0003-1971-5551,sname='Vukadinović']{Dušan~Vukadinović} \affiliation{Institut für Physik, Universität Graz, Universitätsplatz 5, 8010 Graz, Austria}\affiliation{Max-Planck-Institut für Sonnensystemforschung, Justus-von-Liebig-Weg 3, 37077 Göttingen, Germany}\email{dusan.vukadinovic@uni-graz.at}


\begin{abstract}
Flares are caused by an abrupt release of magnetic energy in the solar atmosphere. Plasma heated to well over 10\,MK filling the post-flare corona originates from a rapid heating and ablation of the cooler chromospheric material. This chromospheric evaporation is thought to be facilitated primarily by nonthermal electrons impinging on to the lower atmosphere. Questions on when and where in the chromosphere these upflows originate, however, are not fully resolved. Here we report on unprecedented high-resolution observations of an M-class flare recorded by the Sunrise Ultraviolet Spectropolarimeter and Imager on board the balloon-borne \sunrise{} observatory, that reveal highly structured upflows on spatial scales of $\sim$100\,km originating deep in the chromosphere. The flows even precede the onset of nonthermal electrons by about 10\,minutes and last through the impulsive phase of the flare. Our observations shed new light on the lower atmospheric heating and mass circulation in flares that are challenging to reconcile with the standard solar flare model.
\end{abstract}


\section{Introduction}

Flares are intense disturbances  in the solar atmosphere, which are driven by a rapid release of magnetic energy up to 10$^{33}$\,erg on timescales of minutes to an hour \citep[][]{2011LRSP....8....6S,2020Sci...367..278F}. In the standard solar flare model \citep[][]{1964NASSP..50..451C,1966Natur.211..695S,1974SoPh...34..323H,1976SoPh...50...85K}, magnetic field lines reconnect at a coronal current sheet. The magnetic energy thus released is transferred to plasma heating, particle acceleration and coronal mass ejections \citep[][]{2002A&ARv..10..313P}. Part of the released energy in the corona is transported downward and is deposited in the chromosphere, leading to bright ultraviolet (UV) radiation, X-ray bremsstrahlung and chromospheric ablation (also termed evaporation), resulting in the characteristic post-flare arcade filled with gas heated to a few MK to about 20\,MK \citep[][]{2024ARA&A..62..437F}.

Indeed, upflows of heated plasma are observed to originate at flare ribbons, the locations of energy deposition in the chromosphere and transition region \citep[][]{2009ApJ...699..968M,2014ApJ...797L..14T,2015ApJ...811..139T,2018ApJ...864...63P}. But the processes that govern this intense chromospheric heating and plasma upflows into the corona are debated. One often considered scenario is that the nonthermal electrons accelerated in the corona, in the course of being stopped and thermalized in the denser lower atmosphere, drive this process \citep[][]{1985ApJ...289..414F,2020ApJ...895....6G,2024ApJ...970...21K}. At the same time, heat flux transported from the corona to the chromosphere through thermal conduction is another widely invoked mechanism to explain chromospheric upflows in a flare \citep[][]{1988ApJ...329..456Z,2009A&A...498..891B,2022A&A...657A..51L,2024A&A...684A.171D}. A third process involves the energy transport through the propagation of large-scale Alfv\'en waves pulses from the corona to the lower atmosphere during the impulsive phase of the flare \citep[][]{2008ApJ...675.1645F,2013ApJ...765...81R,2016ApJ...818L..20R}. In addition to the evaporation flows during impulsive phase, it is also known that some transition region and coronal spectral lines show blueshifts (indicative of upflows) already during the precursor and early phases of flares \citep[][]{2004ApJ...613..580B,2016ApJ...823...41D}.

But how deep into the chromosphere can such flows (during both preflare and impulsive phases) be traced remains an open question. Addressing this will provide important constraints to flare energy transport models as higher energy nonthermal electrons are required to penetrate deeper into the chromosphere. Owing to small spatial and temporal scales involved in the magnetic energy release, however, high-resolution imaging spectroscopic data of the solar chromosphere are crucial in assessing where and how evaporation flows originate in a flare. In this study we present unprecedented UV spectroscopic observations capturing the onset of chromospheric evaporation in a solar flare, recorded by the \sunrise{} balloon-bourne observatory \citep[][]{2010ApJ...723L.127S,2017ApJS..229....2S,Solanki_2026,2011SoPh..268....1B,2025SoPh..300...75K}, and discuss implications for the energy transport scenarios in the standard solar flare model.

\section{Observations\label{sec:obs}}

\begin{figure*}
 \begin{center}
   \includegraphics[width=\textwidth]{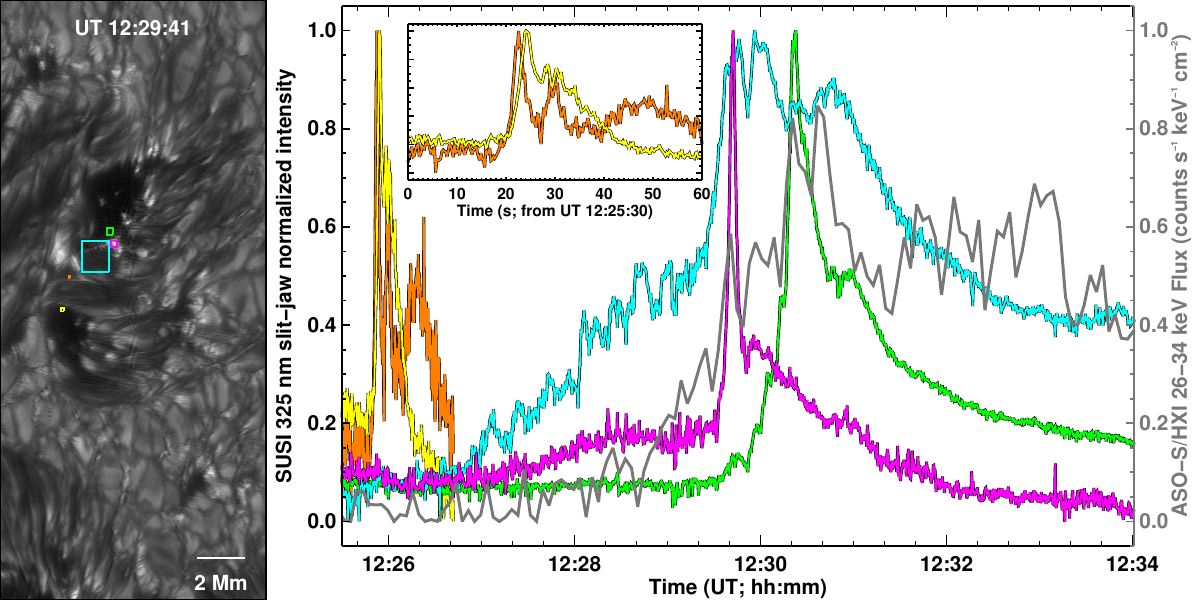}
   \caption{Lower atmospheric small-scale response in the pre- and impulsive phases of the flare. Left: SUSI slit-jaw image covering a footpoint region of the flare. The various colored boxes point to locations of flare ribbons. Right: Normalized emission as a function of time, averaged over the corresponding colored boxes identified in the left panel. The gray curve is the background-subtracted ASO-S/HXI hard X-ray flux in the 26\,keV to 34\,keV energy range. In the inset we show the zoomed-in view of the yellow and orange light curves.\label{fig:lc}}
 \end{center}
\end{figure*}

\begin{figure*}
 \begin{center}
   \includegraphics[width=\textwidth]{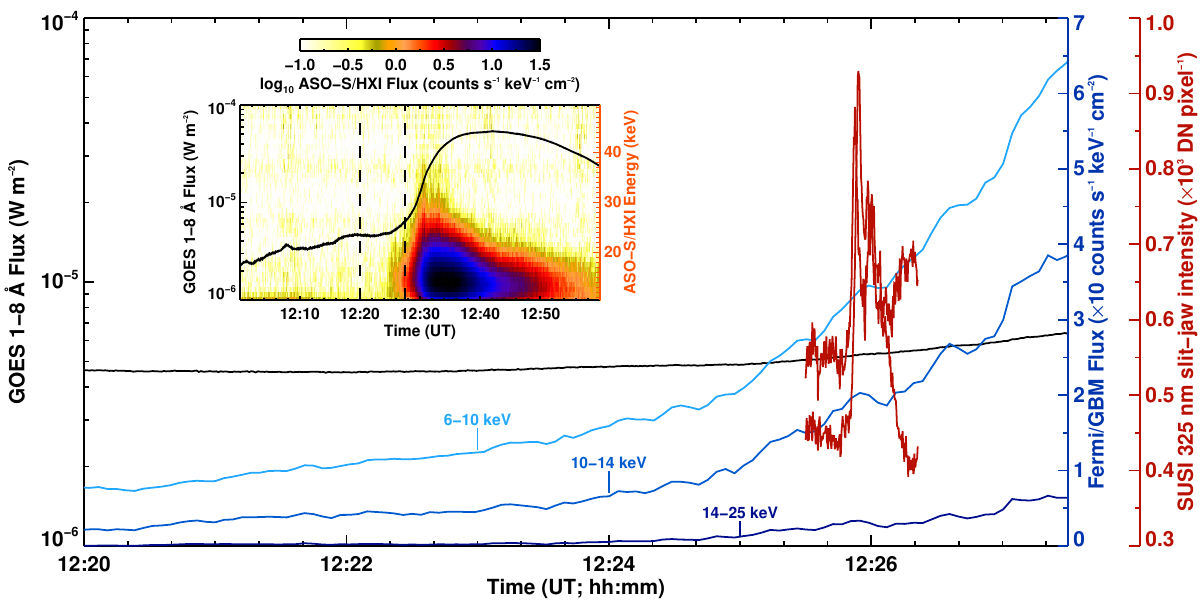}
   \caption{Multi-wavelength signatures in the early phase of the flare. Disk-integrated GOES soft X-ray flux (black) and Fermi/GBM hard X-ray flux in three energy bins (shades of blue) are plotted as a function of time. The maroon-colored light curves are the early flare-related impulses that SUSI captured (same as the yellow and orange curves in Fig.\,\ref{fig:lc}). In the inset, ASO-S/HXI background-subtracted spectrogram (energy vs. time) is shown along with an overlay of the time series of GOES X-ray flux (black). The two vertical dashed lines in the inset mark the temporal boundaries of the main plot. \label{fig:susi_hxi}}
 \end{center}
\end{figure*}

The flare of M5.3 class was observed in the active region with the National Oceanic and Atmospheric Administration (NOAA) number 13738 at Helioprojective Cartesian coordinates of $x\approx+600''$ and $y\approx-150''$ on 2024 July 13 UT\,12:30 \citep[][]{Chitta_2026}. The \sunrise{} data used in the analysis belong to the observation timeline with SUNRISE ID: \texttt{12\_FLAR} \citep[][]{Solanki_2026}. The Sunrise Ultraviolet Spectropolarimeter and Imager \citep[SUSI;][]{2025SoPh..300...65F,2025SoPh..300...58I}, a long-slit scanning spectrograph, was operating in a full-spectral-scan raster mode, sequentially covering 17 different spectral windows in the ultraviolet (UV) wavelength range between about 310\,nm and 400\,nm. The slit-jaw camera on SUSI (SJC), a band-pass imager with a filter having full-width at half maximum of 0.9\,nm centered on 325.407\,nm, provided context images for the raster scans. From UT\,11:44 to UT\,14:33, SUSI observed a portion of the active region, capturing the footpoints of the flare at that location (Fig.\,\ref{fig:lc}). According to the GOES observations, the soft X-ray (SXR) flux started gradually rising around UT\,12:26, peaking around UT\,12:40 (Fig.\,\ref{fig:susi_hxi}). This interval thus falls within the SUSI observing window. 

Each individual raster had a spatial extent of about 5$''$ in the scan direction, and a scan duration of about 500\,s. Here we focus on two particular raster scans, one centered at 325\,nm and the other at 329\,nm, for our analysis. The 325\,nm raster scan covered the period from UT\,12:19 to UT\,12:27, while the 329\,nm raster scan covered the interval from UT\,12:28 to 12:36. Together, they spanned the preflare and the impulsive rise phase of the flare. Taking advantage of the high frame rate of the SUSI detectors, we reconstructed slit-jaw images at an effective cadence of 0.25\,s using a multi-frame blind de-convolution technique \citep[][]{2005SoPh..228..191V}. This high-cadence sequence spans a period of 8\,minutes starting at UT\,12:25:30. The spectral scans are reconstructed with raster step increments at about 0.5\,s cadence. From the first raster we analyze the Fe\,{\sc ii}\,325.59\,nm line, while from the second we target the Fe\,{\sc ii}\,329.58\,nm line. Given the near-seeing-free conditions at which the observations were recorded, the spatial sampling of SUSI spectral scans is diffraction-limited, which is about 0.082$''$ at 325\,nm (equivalent to 59\,km on the solar disk center at 1\,au), making these the highest spatial resolution multi-wavelength UV spectroscopic observations of a flare ever recorded. 

For SUSI spectral analysis, we derived a pseudo-continuum map for each raster, which we computed by averaging the intensity at those wavelength positions where there are no well-defined spectral lines within the corresponding spectral window. This pseudo-continuum map is used to pixel-wise normalize the spectrum. By averaging a portion of SUSI field of view devoid of strong magnetic field concentrations, we computed the average quiet-Sun spectrum, which is normalized to its pseudo-continuum intensity.

Furthermore, we complemented SUSI imaging spectra with the hard X-ray (HXR) data from the Fermi Gamma-ray Burst Monitor \citep[Fermi/GBM;][]{2009ApJ...702..791M}, and the Hard X-ray Imager on board the Advanced Space-based Solar Observatory \citep[ASO-S/HXI;][]{2023SoPh..298...68G,2019RAA....19..160Z,2019RAA....19..163S,2021RAA....21..136C}. The Fermi/GBM data were available only until UT\,12:28, and so we used these to investigate the properties of the preflare phase of this event (details given in Appendix\,\ref{app:hxr}).

\section{Small-scale signatures of the flare in the deep chromosphere\label{sec:deep}}

\begin{figure*}
 \begin{center}
   \includegraphics[width=\textwidth]{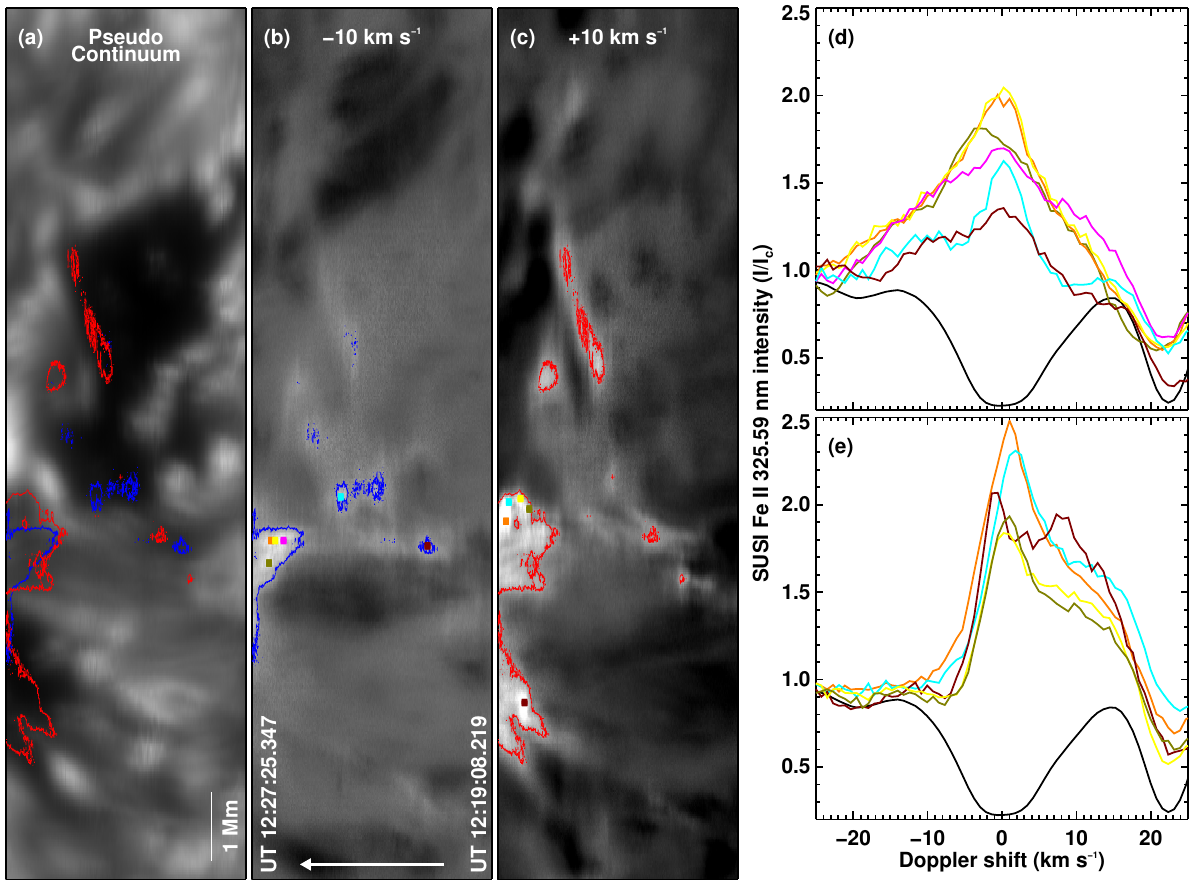}
   \caption{Lower atmospheric imprints of flare ribbons. Pseudo-continuum (a), blue wing map (b), red wing map (c) around the Fe\,{\sc ii} 325.59 nm line are displayed. The blue (red) colored contours enclose regions with enhanced blue (red) wing enhancements. In panel (b) the timestamps mark the beginning and end of the displayed raster, while the scan direction is indicated by the arrow. Sample spectral lines from regions showing blue (d), and red (e) wing enhancements are plotted as a function of Doppler shift. These profiles are normalized to their local continua. Spatial locations of the profiles are marked with the respective colored boxes, separately, in panels (b) and (c). The black curve in both panels is the same, showing the average quiet-Sun spectral profile, normalized to the continuum. \label{fig:spect_325}}
 \end{center}
\end{figure*}

\begin{figure*}
 \begin{center}
   \includegraphics[width=\textwidth]{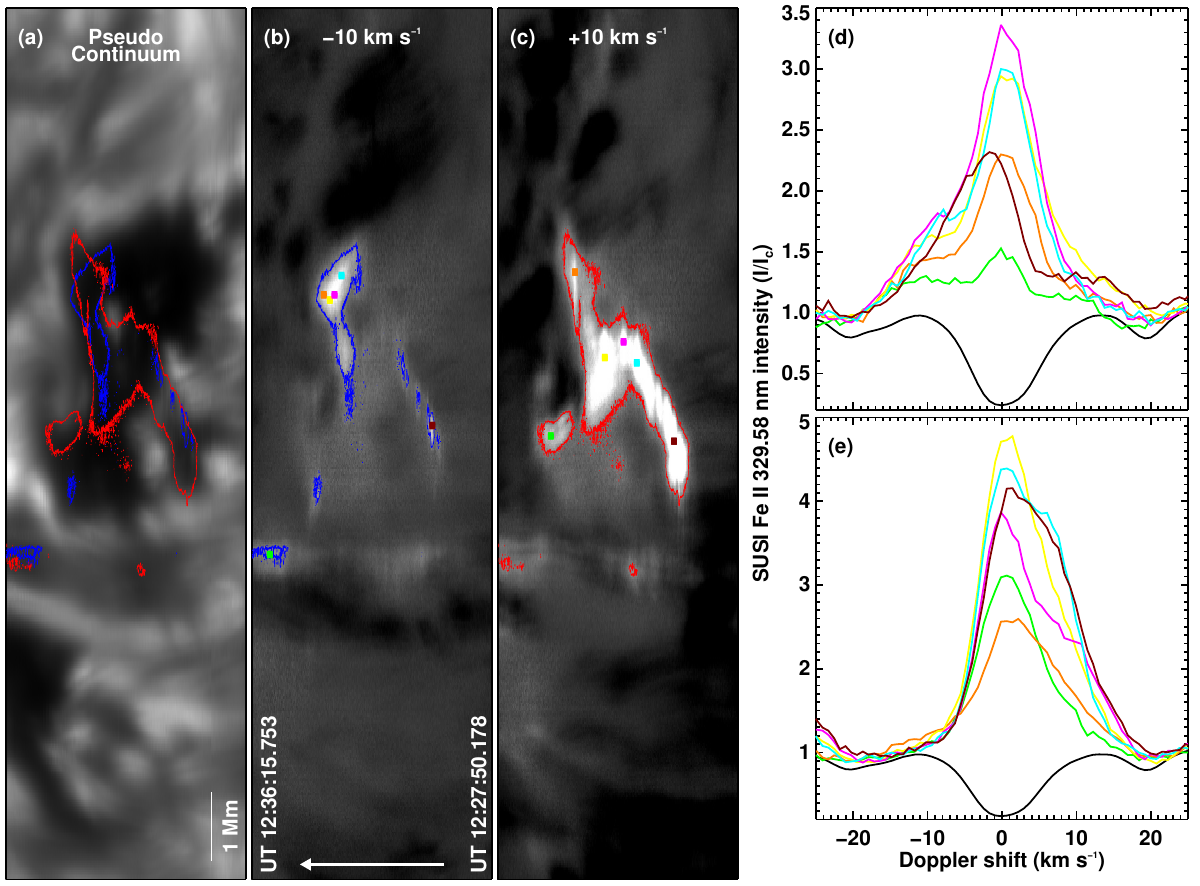}
   \caption{Same as Fig.\,\ref{fig:spect_325} but plotted for the spectral window covering the Fe\,{\sc ii} 329.58 nm line.\label{fig:spect_329}}
 \end{center}
\end{figure*}

\begin{figure*}
 \begin{center}
   \includegraphics[width=0.45\textwidth]{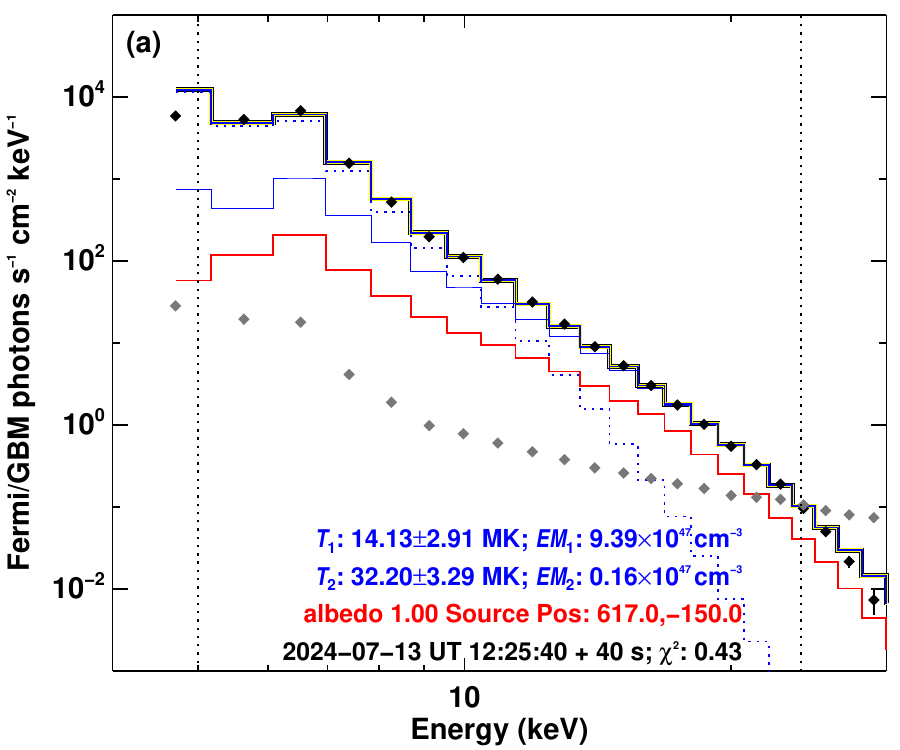}
   \includegraphics[width=0.45\textwidth]{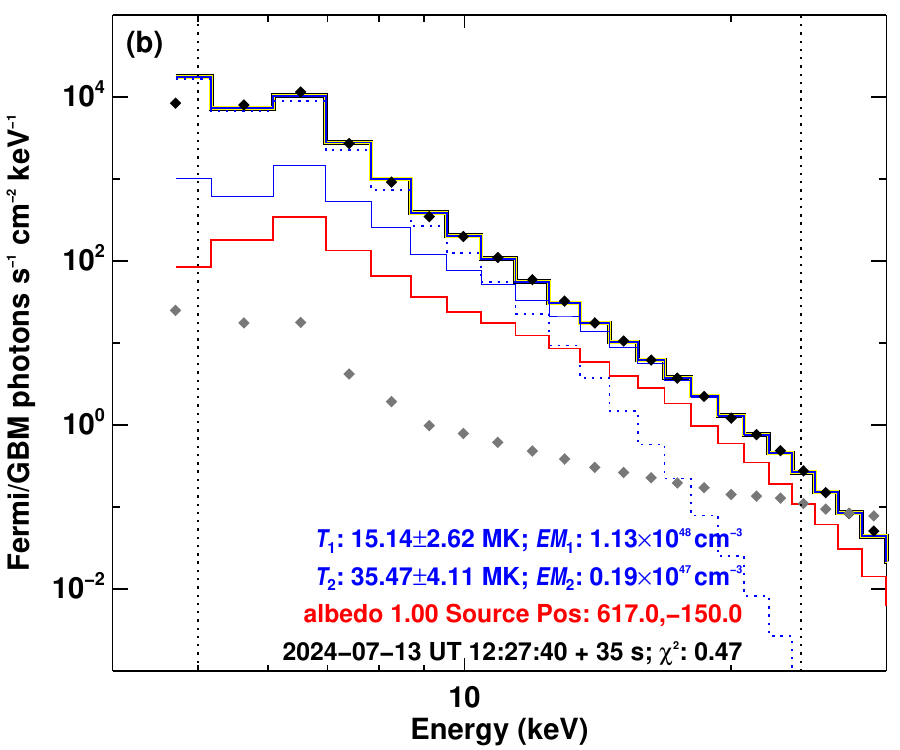}
   \includegraphics[width=0.45\textwidth]{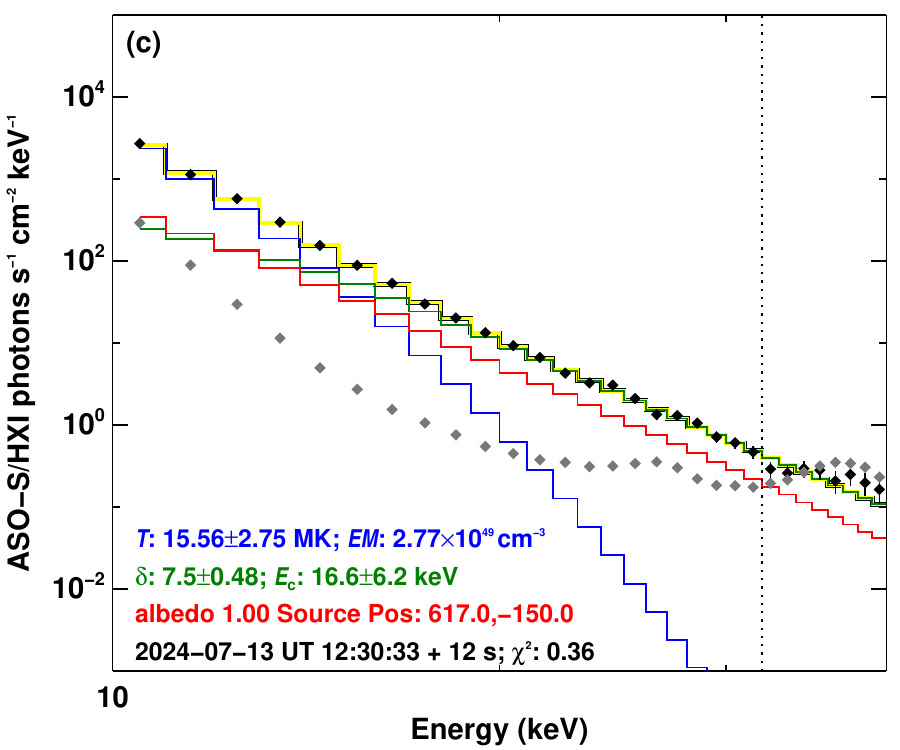}
   \includegraphics[width=0.45\textwidth]{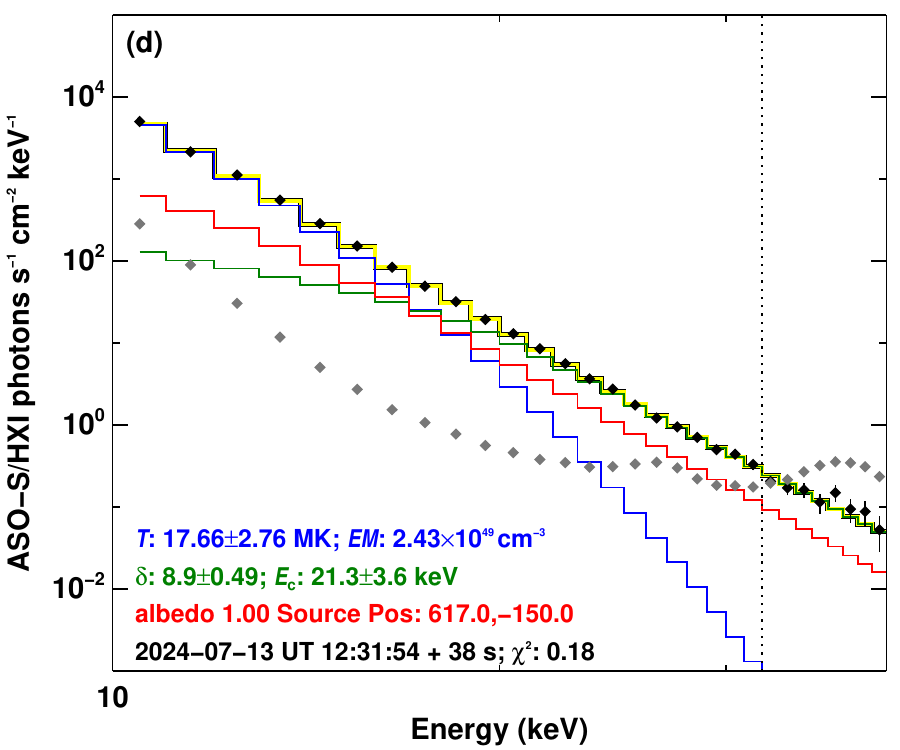}
   \caption{Hard X-ray diagnostics for various time periods during the flare. (a) The background-subtracted Fermi/GBM HXR photon flux as a function of energy, integrated over the specified time period, is plotted with black symbols. The error bars represent 1$\sigma$ uncertainties. The background spectrum is denoted by gray symbols. Blue curve is the two-component isothermal model fitted to the observations with the corresponding fit parameters, temperature $T$, and the emission measures $EM$, listed. Individual components are shown as dotted blue (lower $T$) and solid thin blue (higher $T$) curves. Albedo, with the source position defined in helioprojective-cartesian coordinates, is plotted in red. Yellow curve is the total fit. The goodness of the fit ($\chi^2$) is also quoted. The dotted vertical lines denote the energy range considered for spectral fitting (b) same as panel a but plotted for a different time interval. In panels (a) and (b), blue and yellow curves nearly overlap. (c) The black and gray symbols have the same meaning as in panel a but are obtained from the ASO-S/HXI observations. Isothermal (blue) and thick-target (green) models fitted to the data and the source albedo (red), and the total fit (yellow) are over-plotted. Various fit parameters ($T$, $EM$, spectral index $\delta$, and the low-energy cutoff $E_{\rm c}$) are listed. (d) Same as panel c, but plotted for a different time period.\label{fig:hxr}}
 \end{center}
\end{figure*}

According to the ASO-S/HXI observations, the high-energy HXR signal above 26\,keV started increasing after UT\,12:29, which we define here as the impulsive onset epoch of the flare (Fig.\,\ref{fig:lc}). This is preceded by a preflare phase, with detectable enhancements at the sites of four flare ribbons in the lower chromospheric Mg\,{\sc i}\,b$_2$\,517.3\,nm line for at least 10\,minutes prior to the impulsive onset \citep[][]{Chitta_2026}. The SUSI/SJC captured one of these ribbon sites overlying a group of pores. The SJC primarily covers various absorption lines and the local continuum through its filter. Accordingly, the images predominantly show the continuum features such as the granulation in the field of view. However, at the ribbon brightening locations, there is also contribution from the emission due to the Fe\,{\sc ii}\,325.59\,nm line that falls within the passband. Under quiet-Sun conditions, both the Fe\,{\sc ii} lines considered here form deep in the chromosphere between 0.5\,Mm and 1\,Mm above the solar surface (Appendix\,\ref{app:rh}). Light curves from the various ribbon sources exhibit rapid intensity fluctuations on timescales of less than 10\,s. The flare ribbons remain discernible and show distinct intensity variations relative to their surrounding regions in the SJC images. There are multiple HXR peaks starting around UT\,12:29:40 that temporally coincide with emission peaks from different SUSI/SJC ribbon sources. Moreover, we found cases with spatial scales of the sources reaching the resolution limit of SUSI (for example, yellow and orange boxes overlaid on the SUSI/SJC map in Fig.\,\ref{fig:lc}).

\subsection{Spectroscopic imprints in the deep chromosphere\label{sec:spect}}

Are there any spectral signatures in concurrence with these deep chromospheric ribbon features? To shed light on this question, we now analyze the SUSI raster scans. The pseudo-continuum map reveals typical convective patterns without any indication of flare related disturbances (Fig.\,\ref{fig:spect_325}a); the magnetic context of this flare is provided in \citet{Chitta_2026}. In contrast, consistent with the locations of ribbon brightenings that we observed in the SJC image sequence, SUSI raster maps around the Fe\,{\sc ii}\,325.59\,nm wavelength range show clear intensity enhancements. This extends up to Doppler shifts of $\pm$10\,km\,s$^{-1}$ from the line core. The brightenings are observed both in the interior of the pore and the adjacent structures similar to those seen in the penumbra (panels b and c). The intensity enhancements in the blue- and red-wing maps (outlined by blue and red contours) spatially overlap in some regions. 

Spectral profiles from these preflare ribbon sites show that the Fe\,{\sc ii}\,325.59\,nm line core goes into emission, whereas the typical quiet-Sun spectrum is in absorption (panels d and e of Fig.\,\ref{fig:spect_325}). A blue wing enhancement is seen as a distinct component at about $-10$\,km\,s$^{-1}$ Doppler shift from the line core. Similarly, the profiles with a red wing enhancement exhibit a secondary component at about $+10$\,km\,s$^{-1}$ Doppler shift from the line core. Temporally, some of the regions are observed to exhibit Fe\,{\sc ii} core emission and Doppler shifts (both blue and red) as early as UT\,12:21:24, some 5--10\,minutes prior to the rise of both lower-energy and high-energy HXR enhancements (e.g., maroon colored spectral profile in Fig.\,\ref{fig:spect_325}d).

The raster scan during the impulsive rise phase of the flare, which includes the second Fe\,{\sc ii} line at 329.58\,nm, is displayed in Fig.\,\ref{fig:spect_329}. During this time frame, the flare ribbon regions showing the Fe\,{\sc ii} line blue- and red-wing enhancements are mainly clustered in the pore region. We found that the sites exhibiting blue-wing enhancements are typically weaker in intensity and are structured in narrow lanes while the red-wing features display overall stronger intensity including in the line core, and are spread over a larger region. This behavior remains even when the spectral profiles are not normalized to their local continuum. Another interesting point to note is that the blue-shifted sites are located closer to the outer boundaries of the red-shifted sites. Similar to the preflare phase, we observed clear Doppler components at $\pm$10\,km\,s$^{-1}$ adjacent to a nearly stationary line core. 

\subsection{Nature of the hard X-ray emission\label{sec:hxr}}

While there was no clear high-energy HXR signal during the preflare phase, based on the Fermi/GBM observations, we found that the lower-energy component of the HXR emission (particularly below 14\,keV) was detectable, which exhibited a gradual rise for about 10\,minutes before the impulsive onset. This covers the time period during which we detected early ribbon emission in the SUSI/SJC data Fig.\,\ref{fig:susi_hxi}.  Despite being a moderate M-class event, the bulk of the HXR signal remained below 32\,keV throughout the flaring period (inset in Fig.\,\ref{fig:susi_hxi}).

To understand the nature of the HXR emission at different epochs during this flare, we employed the Object Spectral Executive software \citep[OSPEX;][]{2002SoPh..210..165S} to fit the observed spectra with various models. Background-subtracted X-ray photon flux as a function of energy for the four identified intervals are shown in Fig.\,\ref{fig:hxr}. Given the location of the flare on the solar disk, we included the effect of photospheric Compton back-scattering of X-rays, albedo \citep[][]{2006A&A...446.1157K}, when fitting the observed HXR spectra. During the pre-flare phase (panels a and b), the Fermi HXR spectra are consistent with a two-component thermal bremsstrahlung radiation model. The temperature ($T$) and the emission measure ($EM$) of the dominant component from the model are about 15\,MK and 10$^{48}$\,cm$^{-3}$. These values are consistent with $T$ and $EM$ values retrieved from the GOES SXR signal. The inferred $EM$ of the weaker component is a factor of 60 lower than its dominant counterpart, while the plasma is super heated to temperatures in excess of 30\,MK. Overall, the $T$ and $EM$ values increase between the two phases depicted in panels (a) and (b). We add a caution that the two-component model that we employed here is an approximation because plasma is generally multi-thermal in the corona and will not be limited to these two distinct isothermal states.

The HXR photon flux as observed by the ASO-S/HXI during the impulsive phase of the flare is displayed in the lower two panels of Fig.\,\ref{fig:hxr}. The first of these periods covers a prominent short-living HXR burst that we detected around UT\,12:30:40, while the second one corresponds to the interval when the bulk of the HXR signal peaked around UT\,12:32. The spectrum in both cases are fitted with a combination of an isothermal and a thick-target bremsstrahlung radiation model \citep[][]{2011SSRv..159..107H}. The fitted isothermal component during these periods around the impulsive phase yields temperatures comparable to that of the preflare dominant component, while the $EM$ increases by an order of magnitude. The nonthermal components of the models indicate that the electron spectral index is rather soft, with values ranging between 7 and 9, while the low-energy cutoff is in the range of 10\,keV to 25\,keV. Our analysis thus reveals that at the impulsive phase, the thick-target bremsstrahlung is produced predominantly by low energy electrons. Interestingly, the phase when bulk of the HXR signal peaked is also consistent with a two-component thermal bremsstrahlung model (Appendix\,\ref{app:hxr}). The $T$ and $EM$ values derived in this case follow the increasing trend that we observed during the preflare phase, with the dominant component reaching temperatures above 18\,MK and the minor super-heated plasma emitting at about 45\,MK. This suggests that at the HXR peak, the observed flare and the associated bremsstrahlung radiation could actually be thermal in nature. 

\section{Implications for the energy transport in solar flares\label{sec:disc}}

The temporal evolution of the ribbon emission appears to show distinct properties based on the location (Fig.\,\ref{fig:lc}). For instance, the cyan and green light curves obtained from the pore regions show that the intensity increase is modulated by low-level fluctuations. The other light curves obtained from or close to the edges of the pore, however, display a monotonic intensity enhancement. In case of the cyan curve obtained from a larger region within a pore, the intensity rise is gradual and the fluctuations are noticeable even before the impulsive onset of the flare, whereas the green curve (from a smaller region within the pore) exhibits such fluctuations almost concurrent with the impulsive rise of the HXR emission. It is generally known that flares exhibit quasi periodic pulsations (QPPs) in the electromagnetic spectrum \citep[][]{2016SoPh..291.3143V}. If the fluctuations that we observed are a deep chromospheric manifestation of such QPPs, then their timescales of 5--10\,s are comparable to QPPs detected in the high-cadence observations of the Balmer continuum \citep[][]{2025ApJ...983L..41S}, upper chromosphere and transition region \citep[][]{2026NatAs..10...54A}, and similarly comparable to modulations exhibited by ribbon threads that extend into the corona \citep[][]{2026A&A...705A.113C}. But the observed temporal disparity between different regions that are in proximity to each other is interesting to note. This behavior is further reflected by individual HXR fluctuations, during the impulsive phase of the flare, being concurrent with lower atmospheric intensity emissions at different sites. This could point to small-scale intricacies of flare energy transport to the lower atmosphere.  

To our knowledge, there have been no previous observations of flares with the Fe\,{\sc ii} lines considered here. Ribbon brightenings in the lower atmosphere preceded the impulsive phase by about 10\,minutes \citep[][]{Chitta_2026}. Such preflare UV brightenings are a general feature of a solar flare \citep[][]{2001ApJ...560L..87W}. Here we spectroscopically detected that the preflare brightenings show signatures of highly structured up- and downflows, on small spatial scales, deep in the chromosphere. Intriguingly, we also found that a narrow region at the edge of the pore carrying signatures of spectral red-shifted components in the preflare phase (Fig.\,\ref{fig:spect_325}c) turned to exhibit line profiles with blue-shifted component in the impulsive phase (Fig.\,\ref{fig:spect_329}b).

The Fe\,{\sc ii} intensity at locations with upflows is fainter compared to that in the downflow regions. This is likely because the mass flux carried by the upflows will be rapidly heated to higher coronal temperatures and thus not all of it is sampled by the Fe\,{\sc ii} lines. The existence of plasma emission above 30\,MK that we observed in this flare could result from a part of this upflowing plasma getting heated to such high temperatures. Such a hot plasma in the preflare phase could explain the existence of super-heated plasma with temperatures above 30\,MK that we observed in this event. Although, such high temperatures are thought to occur almost exclusively in X-class flares, which are more energetic than the one analyzed here \citep[][]{2014ApJ...781...43C}. 

As such, observations of other Fe\,{\sc ii} lines, particularly the one at 281.445\,nm, typically show strong red wing asymmetries and secondary components up to $+$40\,km\,s$^{-1}$ from the line core in flares, indicative of downflowing of the heated chromospheric material, which are reproduced by the one-dimensional (1D) radiation hydrodynamic models of flares \citep[][]{2017ApJ...836...12K,2020ApJ...895....6G}. In this direction, to understand the nature of Fe\,{\sc ii} emission, we employed a grid of publicly available 1D models of flares \citep[][]{2023A&A...673A.150C}, whose electron beam parameters are comparable to the observed values (Fig.\,\ref{fig:hxr}; details presented in Appendix\,\ref{app:rh}). While the synthesized Fe\,{\sc ii} lines from these flare experiments do show line core emission, they do not lead to the $\pm$10\,km\,s$^{-1}$ Doppler-shifted components that we observe. For comparison, the models presented in \citet{2017ApJ...836...12K} and \citet{2020ApJ...895....6G} have harder nonthermal electron spectra than inferred in our case. A critical question of how the radiative back-warming, that is often invoked to explain the heating around the temperature minimum region \citep[][]{1990ApJ...365..391M}, could explain such Doppler components and spectral transitions between preflare redshifts and impulsive phase blueshifts, remains open.

One possibility is that the observed Fe\,{\sc ii} Doppler components are generated through a localized heating (e.g., lower atmospheric magnetic reconnection). This scenario is similar to that invoked to be responsible for the heating of hot loops in active region cores \citep[][]{2020A&A...644A.130C,Noda_2026}. In this flare, there are at least four ribbon locations and so there is a 25\% chance that SUSI captured the very ribbon feature where the magnetic energy is liberated. But the feasibility of lower atmospheric reconnection in larger flares producing HXR emission with multiple flare ribbons is not well understood. This would imply that lower atmospheric reconnection must accelerate electrons to higher energies \citep[][]{2024A&A...688L...9C}, which ought to be transported to the conjugate footpoints. But how such nonthermal electrons accelerated in the lower atmosphere survive their transport to the conjugate footpoint without being locally thermalized is an open question. As such, present models generally assume that electron energy injection is in the corona.  Energy transport to and from the particular footpoint that SUSI captured is, nevertheless, required to explain the observed lower atmospheric and HXR characteristics of this flare. Alternatively, heat transported via conduction from a coronal energy release site to the chromosphere could drive the upflows \citep[][]{2015ApJ...813..113B}. This can, in principle, explain the signatures of upper chromospheric evaporation inferred using blueshifts of the Fe\,{\sc xxi} line (forming at 10\,MK) even during the preflare phase \citep[][]{2016ApJ...823...41D}. But such a mechanism is inefficient in the lower chromosphere. Concerning energy transported via Alfv\'en waves pulses, questions on the nature of such waves (torsional vs. kink), the flux they carry and their prevalence in the pre-flare phase are not yet constrained by observations. Overall, the observed upflows originating deep in the chromosphere through the impulsive phase, including the preflare phase, are challenging to be explained by the standard solar flare model \citep[][]{1964NASSP..50..451C,1966Natur.211..695S,1974SoPh...34..323H,1976SoPh...50...85K}.

\section{Conclusion}

Using unprecedented high-resolution imaging spectroscopic data obtained by \sunrise{}/SUSI, we observed signatures of flare-driven energy transport in the deep chromosphere to be structured on spatial scales as small as 60--100\,km. These imprints include rapid intensity variations, on timescales of 5--10\,s, of the lower chromospheric Fe\,{\sc ii} line. While the expected condensation downflows are also observed to penetrate into the lower atmosphere, our observations importantly reveal small-scale upflows deep in the chromosphere, either generated in situ or as flare-driven evaporation, with speeds of about 10\,km\,s$^{-1}$ at the sites of flare ribbons. This provides first direct evidence for their early onset already in the preflare phase even in the lower chromosphere, and their persistence through the impulsive rise of the flare hard X-ray emission. Given the small-scale and lower intensity nature of these upflows, compared to  brighter red-shifted condensation, it is highly likely that their detection is missed in previous lower spatial resolution observations. Our high-resolution observations of a flare thus shed new light on the key aspects of a solar flare on small spatial scales. Specifically, the Fe\,{\sc ii} spectral line diagnostics of flares in the previously unexplored wavelength regime, which we presented here, add new constraints to models of flare energy transport. Our observations also point to the importance of having improved observations of hard X-rays, particularly during preflare phase, which are likely limited by the present HXR detectors.

\begin{acknowledgements}
We thank the anonymous referee for constructive suggestions that helped us improve the presentation of the manuscript. L.P.C. acknowledges useful discussions with Adam Kowalski. This project has received funding from the European Research Council (ERC) under the European Union's Horizon Europe research and innovation programme (grant agreement Nos. 10103984 -- project ORIGIN; 101097844 -- project WINSUN). F.A.I. is a member of the “Carrera del Investigador Científico” of CONICET and supported by MPG through the Max Planck Partner Group between MPS and the University of Mendoza, Argentina. The work of W.C. is supported by the NSFC (grant No. 12333010) and the National Key R\&D Program of China (grant No. 2022YFF0503002). ALST acknowledges funding from the Consejer\'ia de Transformaci\'on Econ\'omica, Industria, Conocimiento y Universidades of the Junta de Andaluc\'ia through grant POSTDOC-21-00832. Sunrise III is supported by funding from the Max-Planck-Förderstiftung (Max Planck Foundation), NASA under Grants \#80NSSC18K0934 and \#80NSSC24M0024 (``Heliophysics Low Cost Access to Space'' program), and the ISAS/JAXA Small Mission-of-Opportunity program and JSPS KAKENHI Grant Numbers JP18H05234 and JP23K25916. This research has received financial support from the European Union’s Horizon 2020 research and innovation program under grant agreement No. 824135 (SOLARNET). It has also been funded by the Deutsches Zentrum für Luft- und Raumfahrt e.V. (DLR, grant no. 50 OO 1608). The Spanish contributions have been funded by the Spanish MCIN/AEI/10.13039/501100011033 under projects RTI2018-096886-B-C5, PID2021-125325OB-C5, and PID2024-156066OB-C5, and from ``Center of Excellence Severo Ochoa" awards to IAA-CSIC (SEV-2017-0709, CEX2021-001131-S), all co-funded by ``ERDF A way of making Europe". SDO is the first mission to be launched for NASA's Living With a Star (LWS) Program and the data supplied courtesy of the HMI and AIA consortia. We thank GOES team for making the X-ray data publicly available. ASO-S mission is supported by the Strategic Priority Research Program on Space Science, the Chinese Academy of Sciences, Grant No. XDA15320000.  The research leading to these results has received funding from the European Community’s Seventh Framework Programme (FP7/2007-2013) under grant agreement no. 606862 (F-CHROMA), and from the Research Council of Norway through the Programme for Supercomputing.
\end{acknowledgements}

\appendix

\section{HXR data analysis \label{app:hxr}}

The background-subtracted ASO-S/HXI light curve shown in Fig.\,\ref{fig:lc} was produced by combining the signal in the 26--34\,keV range from the detectors D92, D93, and D94. The background itself was obtained by combining the signal from the HXI background detectors D95, D96, and D99. Both the data and the background were retrieved at 4\,s integration. The same set of detectors and time integration was also used to produce the HXI spectrogram shown in the inset of Fig.\,\ref{fig:susi_hxi}.

For the HXI spectral analysis (Fig.\,\ref{fig:hxr}), the data was obtained by combining the signal from the detectors D92, D93, and D94. The background was obtained also from these detectors, but from the ASO-S orbit 48\,hours prior to the observations. In particular, the background was obtained by integrating the signal in the time range UT\,12:25 to UT\,12:35 on 2024 July 11. For Fermi/GBM spectral analysis, we obtained data from its detector 5 and set the background time to be UT\,11:51 to UT\,11:58 on 2024 July 13. The Fermi/GBM light curves in Fig.\,\ref{fig:susi_hxi} are obtained by subtracting the signal from the three most anti-sunward detectors from the three most sun-ward detectors at the time of observations. 

\begin{figure}
 \begin{center}
   \includegraphics[width=0.45\textwidth]{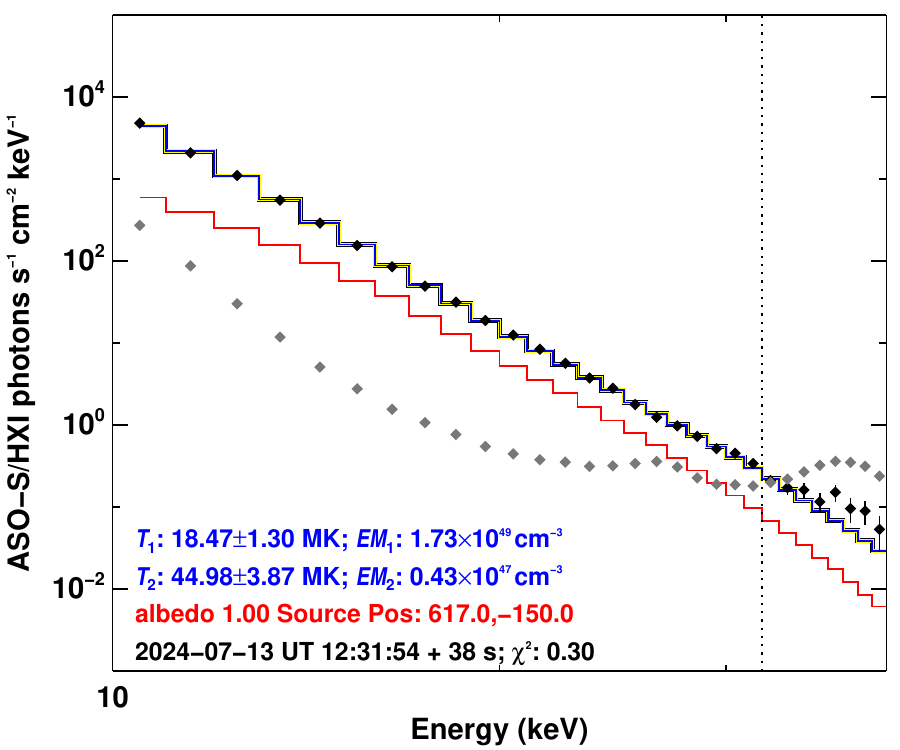}
   \caption{Same as Fig.\,\ref{fig:hxr}d but here the HXR data are fitted with a two-component thermal bremsstrahlung model.\label{fig:hxi}}
 \end{center}
\end{figure}

In the main text we fit the peak HXR phase of the flare with a combination of isothermal and thick-target bremsstrahlung model. In Fig.\,\ref{fig:hxi} we show that the observations are also consistent with a two-component thermal bremsstrahlung radiation model. This suggests that the observed flare is predominantly thermal in nature.

\section{Line synthesis from flare models\label{app:rh}}
The nonthermal electrons  stopped in the dense chromosphere produce the thick-target bremsstrahlung in hard X-rays \citep[][]{2011SSRv..159..107H}. From Fig.\,\ref{fig:hxr}, we note that the HXR spectrum fitted with a thick-target model yields a nonthermal electron spectral index, $\delta$, in the range of 7 and 9, with the low-energy cut-off, $E_{\rm c}$ in between 10\,keV and 25\,keV. During the impulsive phase of the flare (Fig.\,\ref{fig:hxr}c), the total nonthermal power is about $6\times10^{28}$\,erg\,s$^{-1}$. Based on a companion study of the same flare \citep{Chitta_2026}, we estimate that the ribbon emission in the chromosphere is spread over an area of $0.25-1\times10^{18}$\,cm$^{2}$. If this ribbon emission is indeed related to the nonthermal electrons, then the energy flux of the accelerated electrons has a lower limit of $6\times10^{10}$\,erg\,cm$^{-2}$\,s$^{-1}$. In these models, the electron beam has a triangular pulse lasting 20\,s, with flux peaking at 10\,s. The initial condition is a quiet-Sun atmosphere.

To explore how the nonthermal electrons impacted the lower atmosphere in this flare, and to better understand the nature of the observed Fe\,{\sc ii} Doppler components and their formation, we used a publicly available grid of solar flare simulations produced by the F-CHROMA consortium using the one-dimensional radiation hydrodynamics code RADYN \citep[][]{1992ApJ...397L..59C,1995ApJ...440L..29C,1997ApJ...481..500C,2002ApJ...572..626C,2023A&A...673A.150C,1999ApJ...521..906A,2005ApJ...630..573A,2015ApJ...809..104A}. Using the observationally inferred properties of the nonthermal electrons as a guide, we retrieved four flare models ($\delta$ of 7 and 8; $E_{\rm c}$ of 10\,keV and 25\,keV). We selected the models such that they have energy flux contents comparable to the rather large values that we observationally inferred. To this end, we employed the models with a total energy content of $3\times10^{11}$\,erg\,cm$^{-2}$, and a maximum energy flux of $3\times10^{10}$\,erg\,cm$^{-2}$\,s$^{-1}$. 

We synthesized the Fe\,{\sc ii} 325.59\,nm and the Fe\,{\sc ii} 329.58\,nm spectral lines from these flare atmospheres, under the  non-local thermodynamic equilibrium (NLTE) conditions \citep[][]{1980ApJ...241..374C}, using the RH code \citep[][]{2001ApJ...557..389U,2015A&A...574A...3P} and an Fe\,{\sc ii} atom model with 143 levels.\footnote{\url{https://github.com/han-uitenbroek/RH/blob/master/Atoms/FeII_big.atom}} To roughly match the width of the simulated lines with the observed ones, we included the effects of micro-turbulent velocity during the synthesis, which is fixed at 3.5\,km\,s$^{-1}$, as a height and time independent parameter. Additionally, we include opacity fudge factors for the NLTE line synthesis to account for the UV line blanketing and missing opacity \citep[][]{1993A&A...269..509B}.

\begin{figure*}
 \begin{center}
   \includegraphics[width=\textwidth]{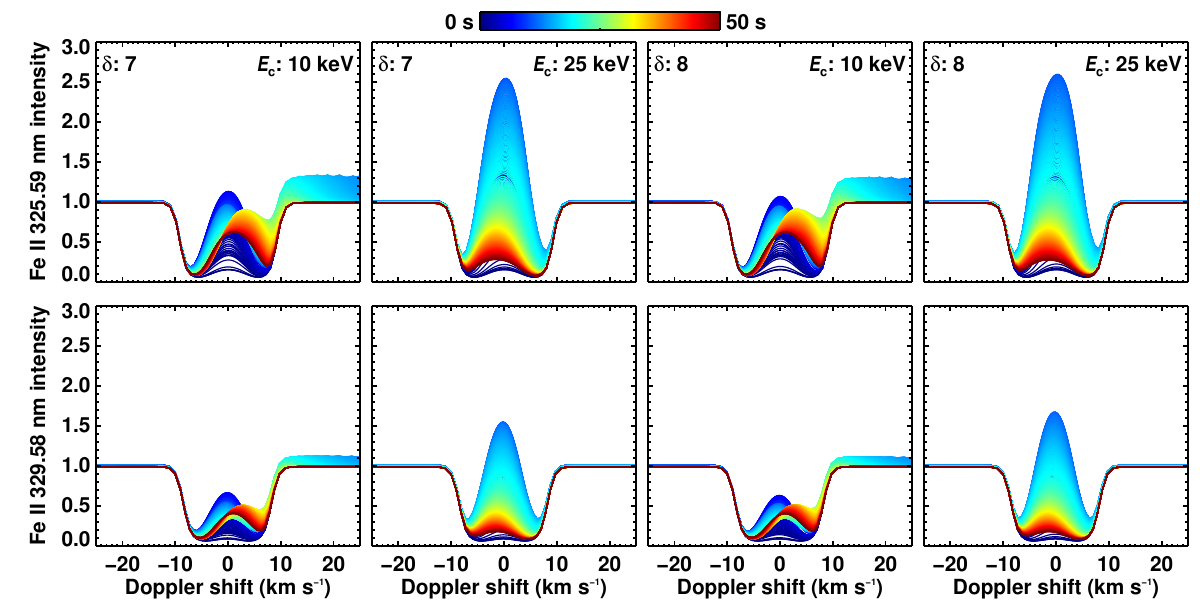}
   \caption{Spectral lines synthesized from RADYN flare atmospheres. Upper panels: Fe\,{\sc ii} 325.59\,nm spectral lines synthesized from a grid of RADYN simulations, normalized to the continuum level. The colors index the temporal progression from 0\,s to 50\,s in the model. Spectral indices ($\delta$) and low-energy cutoffs ($E_{\rm c}$) of the considered models are quoted. Lower panels are for the Fe\,{\sc ii} 329.58\,nm spectral line. \label{fig:rh}}
 \end{center}
\end{figure*}

\begin{figure*}
 \begin{center}
   \includegraphics[width=\textwidth]{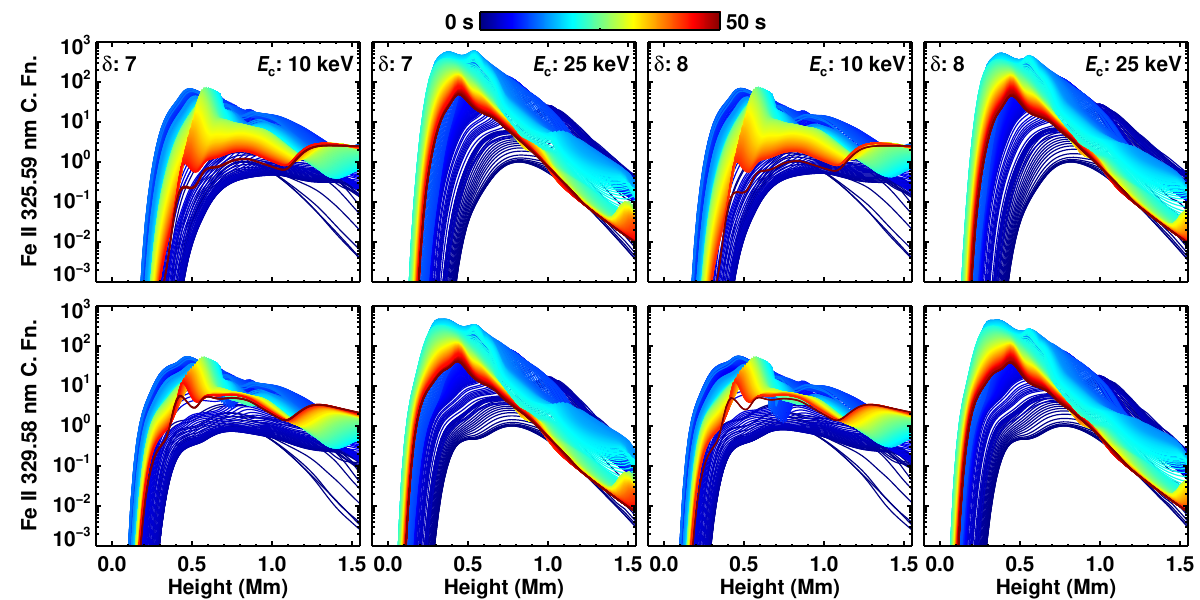}
   \caption{Spectral line formation with height. Contribution functions, normalized to the initial quiet-Sun values, as a function of height are plotted for the grid of RADYN simulations (upper panels: Fe\,{\sc ii} 325.59\,nm line core; lower panels: Fe\,{\sc ii} 329.58\,nm line core). The colors have the same meaning as in Fig.\,\ref{fig:rh}.\label{fig:cfn}}
 \end{center}
\end{figure*}

The synthesized spectral profiles from all the four model atmospheres are displayed in Fig.\,\ref{fig:rh}. These profiles show signatures of emission in the core, with wing asymmetries. The models with the higher value of $E_{\rm c}$ show a more pronounced core emission compared to the lower $E_{\rm c}$ counterparts. This is further elucidated using the contribution functions of the corresponding lines shown in Fig.\,\ref{fig:cfn}. Under the initial conditions of the RADYN simulations, prescribed by a quiet-Sun atmosphere, the contribution function of both the Fe\,{\sc ii} peaks in between 0.5 and 1\,Mm above the nominal $\tau_{500\,\rm{nm}}=1$ height. With time, the line core contribution functions increase in magnitude and begin to peak at lower heights, with a tail towards higher heights. Comparing the two sets of models with differing low-energy cutoffs, 
the ones with $E_{\rm c}=25$\,keV exhibit higher values of contribution functions at lower heights and also a steeper tail at higher altitudes. We found that this difference is mainly caused by variations in the density in the lower atmosphere. 

In these models, the electron beam heating rates peak around 1\,Mm above the solar surface. As shown in Fig.\,\ref{fig:cfn}, the considered Fe\,{\sc ii} lines have their contribution function maxima below these heights, where the effect of radiative back-warming and the associated heating of the lower atmosphere could become important \citep[][]{1990ApJ...365..391M,2018ApJ...857L...2H}. Some RADYN experiments with a harder electron spectrum, albeit at lower fluxes, have led to upflows in the lower atmosphere resulting in a blue-shift of the synthesized Mg\,{\sc ii}\,k3 component \citep[][]{2024ApJ...970...21K}. Similarly, to explain the asymmetric profiles and core enhancements of the Na\,{\sc i}\,D$_1$ during a solar flare, \citet{2016ApJ...832..147K} used RADYN simulations with a harder electron spectrum $\delta=4.2$ and $E_{\rm c}=25$\,keV. Their simulations predicted 2 to 3\,km\,s$^{-1}$ gentle evaporation flows from the deep chromosphere.  However, the existence of such a harder electron spectrum is ruled out in our observations. Overall, these experiments identify that while the nonthermal electrons are able to penetrate into the chromosphere, either their energies or fluxes are not strong enough to lead to substantial blue and red wing enhancements in the deep chromosphere sampled by the Fe\,{\sc ii} lines.

It is possible that a different initial model atmosphere (e.g., a penumbra or a sunspot) which is cooler in the lower part of the atmosphere, and consequently less dense, than the RADYN models we used here could control the properties of electron beam penetration into the chromosphere. This could, for instance, increase or alter the flows in the lower atmosphere with deeper penetrating accelerated electrons \citep[][]{2018ApJ...857L...2H}. A penumbral or sunspot type atmosphere is indeed relevant in our case. But we also noted that the ribbon features and their Doppler components already existed in the preflare phase, at intervals when we inferred plasma super-heated  to temperatures above 30\,MK. This would have its own influence on the atmospheric structure as the excess heat input in the corona would lead to higher density in the upper atmosphere (that is responsible for the GOES SXR emission in the first place). The initial super-heated coronal structure (filled with denser plasma), and the underlying sunspot-like cooler lower atmosphere could play competing roles in the overall flare energy transport, which we have not fully understood.  

\bibliographystyle{aasjournalv7}

\end{document}